\documentstyle[12pt]{report}
\begin{document}
\baselineskip=0.8cm
\ \ \

\vspace{2.0cm}
\begin{center}
{\bf NEAR-INFRARED PHOTOMETRY OF THE CORE-COLLAPSED METAL-POOR GLOBULAR
CLUSTERS NGC5946 AND NGC7099}
\end{center}

\begin{center}
T. J. Davidge$^a$, Department of Geophysics \& Astronomy,

University of British Columbia, Vancouver, BC CANADA V6T 1Z4

and

Canadian Gemini Project Office, Dominion Astrophysical Observatory,

5071 W. Saanich Road, Victoria, BC CANADA V8X 4M6$^b$

{\it email: davidge@dao.nrc.ca}
\end{center}

\vspace{1.0cm}
\noindent{\it To appear in The Astronomical Journal}

\vspace{1.0cm}
\noindent $^a$ Visiting Astronomer, Cerro Tololo Inter-American Observatory,
which is operated by AURA under contract to the National Science Foundation.

\vspace{1.0cm}
\noindent $^b$ Mailing Address

\pagebreak[4]
\begin{center}
ABSTRACT
\end{center}

	Moderately deep near-infrared images are used to investigate the
photometric properties and spatial distribution of bright giants near the
centers of the core-collapsed globular clusters NGC5946 and NGC7099.
The former cluster is located at a low Galactic latitude and is heavily
reddened; nevertheless, the $K, J-K$ color-magnitude diagram (CMD) shows a
well-defined giant branch, with a width $\pm 0.05$ magnitudes in $J-K$. Most of
the observed width is due to differential reddening. The
upper giant branch is relatively steep, and comparisons with
less reddened clusters indicate that (1) NGC5946 must be very
metal-poor, with [Fe/H] $\sim -2$, and (2) $E(B-V) \sim 0.6$. The distance to
NGC5946 derived using the near-infrared brightness of the giant branch tip
is in excellent agreement with that computed from the
$V$ brightness of the HB if $E(B-V) = 0.6$.

	The NGC7099 giant branch terminates $\sim 1$ mag fainter in $K$ than
the expected tip brightness. The integrated $J-K$ color of NGC7099 changes with
radius, such that $\Delta (J-K) /\Delta log(r) = 0.05 - 0.09$, depending on the
technique used to measure colors. Similar radial variations in stellar content
might be expected in NGC5946; however, the integrated $J-K$ color of this
cluster remains constant with radius and, after removing the brightest resolved
stars, a color gradient is seen that is in the {\it opposite} sense to
that in NGC7099. The relative density of bright giant
branch stars also remains constant within 30 arcsec of the center of NGC5946.
Therefore, it is concluded that NGC5946 does not contain stellar content
gradients similar to those seen in NGC7099 and other post core-collapsed
clusters.

\pagebreak[4]
\begin{center}
I. INTRODUCTION
\end{center}

	A large number of globular clusters are located at low Galactic
latitudes, with lines of sight that pass through the disk and bulge of the
Galaxy. Differential reddening and field star contamination complicate
efforts to measure fundamental parameters such as metallicity, reddening,
and distance for many of these objects. Despite these inherent difficulties,
studies of low Galactic latitude globular clusters are potentially
of great importance. For example, if the earliest episodes of star
formation occured near the center of the Galaxy, where the density of
primordial star-forming material may have been highest, then the oldest,
most metal-deficient clusters should be found in the
general vicinity of the Galactic Center. Moreover, it has long been realised
that the spatial distribution of clusters at low Galactic latitudes can be used
to estimate the distance to the Galactic Center (e.g. Shapley 1918; Racine \&
Harris 1989). Finally, it appears that the structural properties of clusters
are related to distance from the Galactic Center and height above
the disk (Chernoff \& Djorgovski 1989), so studies of low latitude clusters may
provide a means of probing the external forces that act on clusters.

	The effects of differential reddening and foreground contamination
can be reduced by observing at wavelengths longward of $1\mu$m. Indeed,
the selective extinction in $K$ is roughly a tenth that in $V$ (e.g. Rieke \&
Lebofsky 1985), while the contrast between the brightest stars and the main
body of the cluster is greatly enhanced in the infrared. The latter point
is important for efforts to study stars in the dense central regions of
clusters, where the relative field star contamination is smallest. In the
present paper, we use near-infrared photometric observations
to investigate the fundamental properties of NGC5946, a
low-latitude globular cluster. NGC5946 is located in a very crowded
field, and is heavily reddened, with $E(B-V)$ in the range $0.5 - 0.6$
(Alcaino et al. 1991, and references therein). The cluster
is very compact, with a light profile indicative of
core collapse (e.g. Trager, King, \& Djorgovski 1995). The only
published color-magnitude diagram (CMD) is that presented by Alcaino et al.
(1991), which is based on moderately deep $B$ and $V$ CCD images. The effects
of differential reddening are readily evident in the width of the red giant
branch (RGB), which is roughly 0.3 mag in $B-V$, and the scatter in the
horizontal branch (HB), which amounts to almost 1 mag in $V$.

	The metallicity of NGC5946 is uncertain. The
integrated spectroscopic properties of NGC5946 are suggestive of [Fe/H] $\sim
-1.4$ (Hesser \& Shawl 1986; Armandroff \& Zinn 1988; Brodie \& Hanes 1986),
while Zinn \& West (1984) conclude that [Fe/H] $\sim -1.37 \pm 0.15$
based on the Q$_{39}$ index. However, Bica \& Pastoriza (1983)
derive [Fe/H] = $-1.56$ from DDO photometry while, based on
the location of the giant branch in the CMD and the reddening infered for the
neighboring cluster NGC5927, Alcaino et al. (1991) argue that
NGC5946 may be very metal-poor.

	Comparisons with clusters at larger projected distances above
the Galactic disk can provide valueable insights into the nature of low
latitude clusters like NGC5946. One possible comparison
cluster is NGC7099. Like NGC5946, the light profile of
NGC7099 contains a central cusp (Trager et al. 1995).
The metallicity of NGC7099 is well determined, with [Fe/H] $\sim
-2.1$ (Minniti et al. 1993, Geisler, Minniti, \& Claria 1992; Zinn \& West
1984). NGC7099 has also been the target of numerous photometric studies
(e.g. Dickens 1972; Alcaino 1978; Alcaino \& Liller 1980; Alcaino \&
Wamsteker 1982; Bolte 1987; Piotto et al. 1987; Buonanno et al. 1988; Richer,
Fahlman, \& VandenBerg 1988; Bolte 1989; Piotto et al. 1990; and Cohen \&
Sleeper 1995), which reveal a rich blue HB population, a conspicuous deficiency
of bright giants near the cluster center, and clear evidence for mass
segregation.

	In the following paper, we discuss near-infrared photometric
observations of NGC5946 and NGC7099. The goals of this study are to (1)
estimate the metallicity, reddening, and distance of NGC5946; and (2) determine
if the radial distribution of stellar content in NGC5946 is like that seen in
NGC7099. Details of the observations are discussed in Section 2, while
near-infrared CMD's, two-color diagrams, and aperture
color measurements for both clusters are presented in
Section 3. In Section 4 the metallicity, distance, and color profile of NGC5946
are examined. A brief summary follows in Section 5.

\begin{center}
II. OBSERVATIONS AND REDUCTIONS
\end{center}

	The NGC5946 images were recorded on the night of UT January 31, 1994
using the Ohio State Infrared Imager/Spectrograph (OSIRIS $-$ Depoy et al.
1993), which was mounted at the Cassegrain focus of the CTIO 1.5 meter
telescope. Each pixel on the 256 x 256 Hg:Cd:Te
array subtended 0.47 arcsec per side. Two fields, one centered on the
cluster core, the other offset 66 arcsec from the cluster center, were observed
through standard Caltech-CTIO filters. A number of standard stars from Casali
\& Hawarden (1992) were observed on this and other nights,
and the transformation coefficients derived from these data
are in excellent agreement with those measured
over the course of many other OSIRIS observing runs
(Depoy 1994, private communication).

	The data for NGC7099 were recorded on the night of UT May 26 1991 using
the University of Hawaii Institute for Astronomy infrared camera (Hodapp,
Rayner, \& Irwin 1992), which was mounted at the Cassegrain focus of the UH 2.2
metre telescope. Like OSIRIS, the detector in the UH camera is a 256 x 256
Hg:Cd:Te array. A single field, centered on the cluster core, was observed
through $K'$ (Wainscoat \& Cowie 1992) and Caltech-CTIO $J$ filters with an
image scale of 0.37 arcsec per pixel. Standard
stars from the list published by Elias et al. (1982) were also observed
during the course of the observing run, and details of the transformation to
the standard system have been discussed by Davidge \& Simons (1994).

	At least three exposures, each offset slightly from the others on the
sky to facilitate the identification and suppression of cosmic rays and bad
pixels (ie. `dithering'), were recorded of each cluster field.
Additional details of the observations, including total
integration times and image quality, are listed in Table 1.

	The data reduction followed standard lines.
Dark frames, recorded immediately following each set of
observations, were subtracted from the raw cluster images.
The results were divided by sky flats, constructed
by median-combining images of various fields recorded on the same night. The
flat-fielded images were registered, sky subtracted, and then combined
by computing the median intensity at each pixel location,
a process that effectively suppresses cosmetic defects and cosmic rays. The
final $K$ images of all three fields are shown in Figures 1, 2, and 3.

\begin{center}
III. PHOTOMETRIC MEASUREMENTS
\end{center}

	The brightnesses of individual stars were measured using the
point-spread function fitting routine ALLSTAR
(Stetson \& Harris 1988), which is part of the DAOPHOT (Stetson
1987) photometry package. The resulting near-infrared measurements were
combined, when possible, with $V$ brightnesses from Alcaino et al. (1991 $-$
NGC5946) and Alcaino \& Liller (1980 $-$ NGC7099). The
brightnesses and colors of those stars with $V, J$, and $K$
photometry are listed in Tables 2 (NGC5946) and 3 (NGC7099).

	The $K, J-K$ diagrams for stars in NGC5946 Fields 1 and 2 with fitting
errors, as computed by ALLSTAR, less than 0.07 magnitude in each bandpass are
shown in Figure 4. The Field 1 $(K, J-K)$ CMD shows a well-defined giant
branch sequence, which can be traced from $K \sim 9.4$ to $K \sim 15$. The
giant branch is moderately wide, with a dispersion of $\pm 0.044$ mag in
$J-K$ when $K \leq 11.75$. The width of
the giant branch is determined by a number of factors, including
photometric errors, differential reddening, star-to-star abundance differences,
and the presence of asymptotic giant branch (AGB) stars. An upper limit to the
smearing introduced by differential reddening can be derived by assuming that
the last two factors are negligible. The photometric uncertainties in $J$ and
$K$ were estimated from numerical simulations, in which scaled versions of the
PSF's constructed for each field were added to the final frames. The
brightnesses of these artificial stars were then measured using DAOPHOT. These
simulations indicate that the dispersion in $J-K$ due to photometric errors
when $K \leq 11.75$ is $\pm 0.015$ magnitudes. Hence, differential reddening
produces no more than $\pm 0.04$ mag widening of the giant branch.

	Because it samples a region with lower mean stellar density and
was recorded with relatively long exposure times, the Field 2 $(K, J-K)$ CMD,
shown in the middle panel of Figure 4, extends almost 1 magnitude fainter
than that of Field 1. In addition to the cluster sequence, a
blue plume of foreground disk stars and blue HB cluster stars is also
evident. Contamination from the blue disk component complicates efforts
to trace the cluster sequence fainter than $K \sim 16$, which is $\sim 2$
magnitudes brighter than the expected location of the main sequence turn-off.

	The $(K, V-K)$ diagram for NGC5946 Field 1 is shown in the left hand
side of Figure 5. The $(K, V-K)$ CMD shows considerable scatter,
as $V-K$ is more susceptible to differential reddening than
$J-K$, although the upper giant branch and the HB are clearly evident.
However, the presence of a blue plume in the Field 2 ($K, J-K$) CMD suggests
that some of the blue HB stars may actually be line-of-sight disk stars.

	The $(K, J-K)$ and $(K, V-K)$ CMD's for NGC7099 are shown in the right
hand panels of Figures 4 and 5, respectively. The giant and HB sequences are
clearly evident in both CMD's. The $(K, J-K)$ CMD of NGC7099
is very similar to that published by Cohen \& Sleeper (1995), although the
current data provides deeper photometric coverage.

	The most luminous giants in NGC5946 have intrinsic $K$ brightnesses
over a magnitude brighter than their counterparts in the central regions of
NGC7099. The brightest giants in NGC5946 occur at $K \sim 9.4$,
while in NGC7099 the brightest giants have $K \sim 10.0$. Using the
reddenings and HB brightnesses listed in Table IV of Armandroff (1989), and
assuming that M$_{V}^{HB} = 0.6$, then the brightest giants in NGC5946
have M$_K = -5.5$, while in the central regions of NGC7099 the brightest
giants have M$_K = -4.4$. Hence, there is a deficiency of bright giants in
the central parts of NGC7099.

	$(K, J-K)$ normal points were derived for NGC5946 and NGC7099
by grouping the observations in $\pm 0.25$ magnitude bins in $K$, and then
computing the mean $J-K$ color in each interval. Only the Field 1 observations
were used for NGC5946. Outlier points were rejected by applying an iterative
$2-\sigma$ rejection algorithm, and the results are listed in Table 4.

	The ($J-H, H-K$) diagram for stars in NGC5946 Field 1 brighter than
$K = 13$ is shown in Figure 6. Data for bright giants
in the metal-poor clusters M13 and M92 (Cohen, Frogel, \& Persson 1978), as
well as the sequence defined by solar neighborhood giants (Bessell
\& Brett 1988), are also shown for comparison. The M13, M92, and solar
neighborhood data have been shifted
using the reddening law of Rieke \& Lebofsky (1985) to account for the
color excess of NGC5946, which is assumed to be $E(B-V) = 0.6$
(Section 4). Although the NGC5946 data show considerable scatter,
it is evident that the locus defined by these observations is in
rough agreement with the sequence defined by other metal-poor globular
clusters.

	The ($J-K, V-K$) diagrams for NGC5946 and NGC7099 are shown in Figure
7, along with the corresponding sequences for giants in M92, M13, and the solar
neighborhood. The NGC5946 observations, as well as those for
NGC7099, fall to the left of the solar neighborhood sequence, and are in rough
agreement with the loci defined by M92 and M13. It is evident that a clear
distinction between globular cluster and solar neighborhood giants can be made
on the ($J-K, V-K$) plane, so optical-infrared two-color diagrams should
provide a means of identifying non-cluster stars in crowded environments.

	There is considerable evidence that dynamical evolution introduces
radial variations in stellar content (e.g. Djorgovski \& Piotto 1992),
and color gradients are one signature of this effect. Using moderately
deep images of NGC4147, Davidge (1992) demonstrated that radial color
variations in globular clusters can be detected at wavelengths longward of
$1\mu$m. Therefore, near-infrared color gradients
should be evident in NGC5946 and NGC7099 $-$ is this the case?
To answer this question, two different methods were used to measure the
near-infrared cluster color profiles. First, direct aperture
measurements were made using the PHOT routine in DAOPHOT. Although this
technique has the benefit of computational simplicity, colors derived in this
manner are susceptible to stochastic variations in the spatial distribution of
the brightest giants, an effect which will be exacerbated in the near-infrared.
Consequently, a second set of color profiles were computed using the technique
described by Piotto et al. (1988), in which each annulus is divided into eight
azimuthal segments, and the median of the various segment colors is adopted as
that for the annulus.

	The resulting measurements, especially those for small apertures, are
sensitive to the adopted location of the cluster centers, which
were determined in the current study by locating the symmetry
points of the cluster light distributions about the observed $x$ and $y$ axes.
The results may be biased by stochastic effects in the distribution of bright
giants. To remove this potential source of error, cluster centers
were determined from $J$ images that were smoothed with a running boxcar
median filter to suppress bright stars. The centroiding accuracies achieved
with this technique are on the order $\pm 0.5$ arcsec.

	The color measurements, which are summarised in Tables 5 (NGC5946) and
6 (NGC7099), are plotted in Figure 8. The errors listed in Tables 5 and 6
reflect uncertainties due to cluster centering, which is assumed to be $\pm
0.5$ arcsec, and the $K$ sky level. The former are most significant
at small radii, while the latter dominate at large radii. The
colors derived using the Piotto et al. (1988) technique are systematically
bluer than those computed from direct aperture measurements.
This result is not unexpected, as the median filter rejects those
areas containing the brightest (and reddest) giants.

	The direct aperture measurements reveal a color gradient in NGC7099,
and the amplitude of this gradient does not change significantly when colors
are measured using the Piotto et al. (1988) technique. A least squares linear
fit to the data in Table 6 indicates that $\Delta
(J-K) / \Delta log(r) = 0.09 \pm 0.02$ (aperture
measurements) and $0.05 \pm 0.01$ (segmental median measurements). The quoted
uncertainties reflect only the scatter in the data points for the adopted
cluster centers. To estimate the uncertainty in the slopes due to centering
errors, aperture measurements were made with the cluster center shifted by
$\pm 0.5$ arcsec along each axis, and the resulting slopes have a standard
deviation of 0.02 units. As for NGC5946, least squares linear fits to the
measurements for this cluster indicate that $\Delta (J-K) / \Delta log(r) =
-0.01 \pm 0.04$ (aperture measurements) and $-0.13 \pm 0.01$ (segmental
median measurements). The uncertainties in these slopes due to
centering errors is comparable to that in NGC7099. The measurements made with
the Piotto et al. (1988) technique suggest that a significant color
gradient is seen in NGC5946; however, the sense of this gradient is opposite to
that in NGC7099.

\begin{center}
IV. COMPARISONS WITH OTHER CLUSTERS
\end{center}

\noindent{\it 4.1: RGB Morphology}

\vspace{0.3cm}
	The slope of the upper giant branch is sensitive to metallicity, in the
sense that metal-poor systems have the steepest sequences. The use of giant
branch morphology to determine metallicity requires a well-defined CMD and
knowledge of cluster distance $-$ quantities which, for many clusters at
low latitudes, can only be obtained in the infrared. In the following Section,
the near-infrared CMD of NGC5946 is used to test the suggestion made by Alcaino
et al. (1991) that this cluster is more metal-poor than [Fe/H] $\sim -1.4$.

	Comparison sequences were constructed from
near-infrared photometric measurements of bright giants in
clusters with well-determined metallicities and reddenings,
using data tabulated by Cohen et al. (1978) and Frogel, Persson, \& Cohen
(1983). A very metal-poor (ie. [Fe/H] $\sim -2$)
giant branch sequence was constructed from observations
of stars in the clusters NGC 4590 (M68), 5024 (M53), 6341 (M92), and 7078
(M15), while a corresponding moderately metal-poor
(ie. [Fe/H] $\sim -1.3$) sequence was derived from
the clusters NGC 362, 1261, 1851, and 2808. The individual cluster sequences
were shifted to a common distance and reddening using HB
brightnesses and $E(B-V)$ values listed in Table IV of Armandroff (1989);
the reddening curve of Rieke \& Lebofsky (1985) was
used to derive $E(J-K)$ from the tabulated $E(B-V)$ values.
A mean locus was then fit by hand to the composite CMD's.

	The two standard sequences are compared with the NGC5946 normal points
in Figure 9. These comparisons were made using the HB brightnesses
listed by Armandroff (1989), while the mean reddening of NGC5946 was allowed
to vary such that each sequence matched the
normal point color at $K = 12$. These comparisons reveal that if
NGC5946 is very metal-poor then $E(B-V) \sim 0.60$, while if it is only
moderately metal-poor then $E(B-V) \sim 0.55$.
It is apparent from Figure 9 that the [Fe/H] $= -2$ sequence is best able
to match the NGC5946 giant branch. Therefore,
the claim made by Alcaino et al. (1991) that NGC5946 is more metal-poor
than [Fe/H] $\sim -1.3$ is supported by these data.

	The integrated near-infrared colors of NGC5946
predict reddenings similar to those derived above.
Using the near-infrared colors summarized in Table 5A of Brodie \& Huchra
(1990), if NGC5946 is very metal-poor then the unreddened color should
be $(J-K)_0 = 0.58$. Hence, if $J-K = 0.91$ (Section 3),
then $E(J-K) = 0.32$, and $E(B-V) = 0.62$, based on the reddening
curve of Rieke \& Lebofsky (1985). On the other hand, if [Fe/H] $= -1.3$ then
the intrinsic color for NGC5946 should be $(J-K)_0 = 0.66$, so that
$E(J-K) = 0.25$ and $E(B-V) = 0.48$.

	The [Fe/H] $= -1.3$ and $-2.0$ sequences are compared with the NGC7099
normal points in Figure 10. The deficiency of bright giants noted in previous
studies of this cluster is readily apparent in Figure 10, and this complicates
efforts to estimate metallicity from giant branch morphology, as it is the
region near the RGB-tip that contains the most information concerning [Fe/H].
Nevertheless, based largely on the brightest normal point, the most metal-poor
sequence gives the best match to the observations, as expected based
on existing metallicity estimates (Section 1).

\vspace{0.5cm}
\noindent{\it 4.2: The Distance to NGC5946}
\vspace{0.3cm}

	Contamination from disk stars complicates efforts to measure the
brightness of the HB in low-latitude clusters, while spatial variations in
reddening make corrections for selective absorption uncertain at optical
wavelengths. Both of these factors conspire to make the distances to many
low-latitude clusters, which are usually based on the location of the HB at
optical wavelengths, highly uncertain. These errors in distance propagate into
metallicity estimates derived from the CMD. Hence, it is desireable
to check individual cluster distances using another standard candle.

	How reliable is the HB distance for NGC5946?
Liller (1983) discusses photographic photometry of six RR Lyrae variables
which, based on their location within the tidal radius, are thought to be
cluster members. However, a number of variables have been discovered outside
the tidal radius, so it is likely that at least some of the `cluster' variables
are actually field stars. In fact, the CMD constructed by Alcaino et
al. (1991) shows only a handful of stars in the instability strip, for which
$V \sim 17.2$. The use of such a small sample of stars may produce a
distance susceptible to systematic errors arising from differential reddening.

	The brightness of the RGB-tip can be used to derive independent
distances to globular clusters. The RGB-tip is a potentially powerful distance
indicator since (1) the calibration is well-defined, with a dispersion of $\pm
0.2$ magnitudes (Frogel, Cohen, \& Persson 1983); (2) RGB-tip stars are very
bright, and hence easier to identify and less prone to field
star contamination than, for example, HB stars; and (3) the technique can be
applied in the near-infrared, where the effects of reddening are much smaller
than at optical wavelengths. Nevertheless, the RGB-tip does not provide a
panacea for the cluster distance scale. For example, stochastic effects are
significant for low mass clusters, and this problem can only be avoided by
restricting the procedure to clusters that are moderately bright. Moreover,
some dynamically evolved clusters, such as NGC7099, show a central deficiency
of bright giants, although these objects are identifiable based on the
radial distribution of bright giants (e.g. Djorgovski \& Piotto 1992). Finally,
the presence of asymptotic giant branch (AGB) stars may confuse efforts to
measure the brightness of the RGB-tip. However, it appears
that the AGB does not extend above the RGB-tip in metal-poor clusters (e.g.
Frogel 1983), so contamination from highly-evolved stars should not be an
issue if applied only to clusters with [Fe/H] $\leq -1$.

	In the present study, near-infrared
photometric measurements of bright giants in the
clusters NGC4590, 5024, 6341, and 7078 listed by Cohen et al.
(1978) and Frogel, Persson, \& Cohen (1983) are used to derive an empirical
calibration for the RGB-tip brightness in metal-poor systems. The $K$
brightnesses of the most luminous RGB stars observed in these clusters, based
on a distance scale with M$_{V}^{HB} \sim 0.6$, are summarized in Table 7.
There is reasonable cluster-to-cluster agreement, and these data suggest
that the RGB-tip in very metal-poor clusters is M$_K \sim -5.5 \pm
0.1$, where the uncertainty is the standard error of the mean.
The brightest giants in NGC5946 have $K \sim 9.4$; hence, if $E(B-V) = 0.6$,
then the distance modulus infered from the RGB-tip is 14.7. For comparison, if
V$_{HB} = 17.2$ (Alcaino et al. 1991), $E(B-V) = 0.6$, and M$_{V}^{HB} = 0.6$
then $\mu_0 = 14.7$, in good agreement with that derived from the RGB-tip. This
consistency suggests that the comparisons in Section 4.1 were made using the
correct relative cluster distances, and indicates that the brightness of the
RGB-tip may be useful as a distance indicator for clusters that are heavily
reddened and/or subject to severe foreground star contamination.

\vspace{0.5cm}
\noindent{\it 4.3: Radial Changes in Stellar Content}
\vspace{0.3cm}

	The ratio of bright giants to HB
stars changes with radius in NGC7099 (Alcaino \& Wamsteker
1982; Buonanno et al. 1988; Djorgovski \& Piotto 1992), and
recent observations with the HST have detected a centrally concentrated
population of blue stragglers (Yanny et al. 1994). Given the
evidence for radial changes in stellar content, it is not surprising
that broad-band color gradients at optical wavelengths have been detected (e.g.
Cordoni \& Auriere 1984; Pastoriza et al. 1986; and Piotto, King, \& Djorgovski
1988), and the amplitude of these gradients are similar to those seen in other
dynamically evolved clusters (e.g. Djorgovski et al. 1991).

	The data presented here reveal that the near-infrared color of NGC7099
changes with radius. The rate with which $J-K$ changes with log(r) in NGC7099
is insensitive to the technique used to measure color, suggesting that the
gradients are not caused exclusively by the brightest giant branch
stars, which the Piotto et al. (1988) technique suppresses. The detection of
near-infrared color gradients in NGC7099 and NGC4147 (Davidge 1992) suggests
that similar radial changes in $J-K$ might be expected in NGC5946. It is
therefore interesting that, despite having a post core-collapse morphology, the
current data fail to detect $J-K$ variations in NGC5946 similar to
those in NGC7099 $-$ in fact, NGC5946 appears to contain a gradient
in the {\it opposite sense} to that seen in other post core-collapsed clusters.
NGC5946 is the only post core-collapse cluster studied to date to show this
behaviour (e.g. Djorgovski \& Piotto 1992). It
could be argued that differential reddening may mask an underlying color
gradient. However, this is unlikely as the width of the giant branch indicates
that differential reddening is at most $\pm 0.04$ mag in $J-K$ (Section 3),
and aperture measurements, especially those which utilize median
filtering techniques, will act to suppress reddening differences over
moderately large spatial scales.

	Studies of the spatial distribution of bright giants at near-infrared
wavelengths provide a means of investigating the radial stellar
content of clusters in a manner that is largely independent of
differential reddening. Djorgovski \& Piotto (1992) found an
absence of bright giants within 10 arcsec of the center of NGC7099. The spatial
distributions of bright giants in NGC7099 derived from the current data is
examined in the two right hand panels of Figure 11,
where the CMD's of stars in annuli within the intervals $0 - 10$ and $10 -
19$ arcsec are compared. These intervals have comparable integrated
$K$ brightnesses, so they should contain similar numbers of stars.
In an effort to quantify the spatial distribution of
bright giants, the number of stars detected within 2
magnitudes in $K$ of the RGB-tip, as derived using the calibration in
Section 4.2, were counted. In the particular case of NGC7099, the RGB-tip
should occur at $K = 9.0$, so the number of stars brighter than $K = 11$ in
each radial interval provide the relevant data. It is evident that whereas
4 stars are seen with $K \leq 11$ in the outer annulus, only 1 star with this
brightness is seen in the inner annulus.

	The radial distribution of bright giants in NGC5946 is investigated in
the two left hand panels of Figure 11. Annuli were selected for NGC5946 that
sample the same intervals in linear units as in
NGC7099, and these have inner-outer radii of
$0 - 9$ arcsec and $9 - 16.5$ arcsec after adjusting for the difference in
distance. In both intervals there are 5 stars within 2 magnitudes of the
RGB-tip ($K = 9.5$), so there is no evidence for radial changes in the number
of bright giants. This result provides further evidence that NGC5946 does not
contain population gradients like those in NGC7099, even though it is a post
core-collapse cluster. It would be of interest to survey a larger sample of
dynamically evolved low-latitude globular clusters at infrared wavelengths
to determine what fraction of these objects contain radial gradients in
stellar content like those seen in NGC7099.

\begin{center}
V. SUMMARY
\end{center}

	The results presented in this paper demonstrate that near-infrared
measurements can be used to study the metallicities, distances, and radial
stellar contents of heavily reddened globular clusters.
The specific conclusions of this study are:

\noindent{1)} The upper giant branch of NGC5946 in the $(K, J-K)$ diagram is
relatively steep, indicating that the metallicity is significantly lower than
[Fe/H] $\sim -1.4$. If NGC5946 is as metal-poor
as the giant branch slope indicates, then $E(J-K) =
0.3$, which corresponds to $E(B-V) = 0.6$ using the reddening curve of Rieke
\& Lebofsky (1985).

\vspace{0.3cm}
\noindent{2)} The color and slope of the RGB in the $(K, J-K)$ diagram of
NGC7099 is consistent with the metal-poor nature of this cluster. There is a
deficiency of giants within 2 mag in $K$ of the giant branch tip in the central
regions of this cluster, consistent with earlier observations at optical
wavelengths.

\vspace{0.3cm}
\noindent{3)} The distance modulus of NGC5946 derived from the brightness of
the RGB-tip is 14.7. This estimate, which assumes that M$_{V}^{HB} = 0.6$ in
very metal-poor clusters, is in excellent agreement with that derived from the
visual brightness of the HB if $E(B-V) = 0.6$.

\vspace{0.3cm}
\noindent{4)} NGC7099 contains near-infrared color gradients with slope
$\Delta (J-K) / \Delta log(r) \sim 0.05 - 0.1$, depending on the technique
used to measure color. Both bright giants and fainter stars
appear to be responsible for the observed near-infrared color variations.

\vspace{0.3cm}
\noindent{5)} Although NGC5946 shows a highly concentrated light profile
suggestive of dynamical evolution, neither the integrated $J-K$ color nor the
specific frequency of bright giant branch stars varies with radius out to a
distance of 30 arcsec from the cluster center. If colors are measured using the
technique described by Piotto et al. (1988) then a color gradient is
detected, but the sense of this gradient is opposite to what is seen
in NGC7099 and other post core-collapsed clusters.

\vspace{1.0cm}
	It is a pleasure to thank Doug Simons and Peter Garnavich for
assistance on the nights these data were taken. Sincere thanks are also
extended to the Natural Sciences and Engineering Research Council of
Canada (NSERC) and the National Research Council of Canada (NRC) for
financial and material support.

\pagebreak[4]
\begin{center}
TABLE 1. Observing Log
\end{center}

\begin{center}
\begin{tabular}{llcclc}
\hline\hline
Cluster & Date & Field \# & Filter & Exposure time & Seeing \\
 & (UT) & & & (sec) & (arcsec) \\
\hline
NGC5946 & February 1, 1994 & 1 & $J$ & $3 \times 20$ & 1.2 \\
 & & & $H$ & $3 \times 20$ & 1.2 \\
 & & & $K$ & $6 \times 20$ & 1.2 \\
 & & & & & \\
NGC5946 & February 1, 1994 & 2 & $J$ & $12 \times 100$ & 1.2 \\
 & & & $K$ & $54 \times 50$ & 1.2 \\
 & & & & & \\
NGC7099 & May 26, 1991 & 1 & $J$ & $3 \times 20$ & 1.0 \\
 & & & $K'$ & $3 \times 45$ & 1.0 \\
 & & & & & \\
\hline
\end{tabular}
\end{center}

\pagebreak[4]
\begin{center}
TABLE 2. NGC5946 Stars with $V$, $J$, and $K$ measurements
\end{center}

\begin{center}
\begin{tabular}{rrrr}
\hline\hline
$n^a$ & $K$ & $J-K$ & $V-K$ \\
\hline
  96 &   14.44 &    0.47 &    2.62 \\
 174 &   12.66 &    1.08 &    4.43 \\
 268 &   13.67 &    0.87 &    3.42 \\
 176 &   12.72 &    0.83 &    3.79 \\
 163 &   12.89 &    0.88 &    3.76 \\
 271 &   13.97 &    0.76 &    5.26 \\
 170 &   12.88 &    0.79 &    3.91 \\
 186 &   13.29 &    0.80 &    3.92 \\
 165 &   14.87 &    0.46 &    2.41 \\
 156 &   13.74 &    0.79 &    4.09 \\
 149 &   15.14 &    0.46 &    2.35 \\
 124 &   13.50 &    0.88 &    4.16 \\
  92 &   13.91 &    0.83 &    2.77 \\
  77 &   14.42 &    0.62 &    2.77 \\
  65 &   14.44 &    0.63 &    2.70 \\
  66 &   14.33 &    0.24 &    2.19 \\
  81 &   13.33 &    0.80 &    3.64 \\
 120 &   10.79 &    0.93 &    4.20 \\
 141 &   12.73 &    0.84 &    4.10 \\
 162 &   10.37 &    1.03 &    4.60 \\
 171 &   11.68 &    0.96 &    4.14 \\
 193 &   11.94 &    0.94 &    4.16 \\
 143 &    9.30 &    1.11 &    4.84 \\
 148 &   11.80 &    0.91 &    4.29 \\
 182 &   12.40 &    0.89 &    4.16 \\
\hline
\end{tabular}
\end{center}

\noindent$^a$ Identification number from Alcaino et al. (1991).

\pagebreak[4]
\begin{center}
TABLE 2. (Con't)
\end{center}

\begin{center}
\begin{tabular}{rrrr}
\hline\hline
$n^a$ & $K$ & $J-K$ & $V-K$ \\
\hline
 131 &   11.88 &    0.89 &    4.23 \\
 116 &   11.88 &    0.93 &    4.19 \\
  97 &   10.98 &    0.11 &    0.91 \\
  79 &   11.37 &    0.88 &    4.04 \\
  80 &   11.81 &    0.85 &    3.86 \\
  83 &   12.86 &    0.78 &    4.19 \\
  75 &   11.97 &    0.83 &    3.98 \\
 115 &   10.27 &    0.99 &    4.54 \\
\hline
\end{tabular}
\end{center}

\noindent$^a$ Identification number from Alcaino et al. (1991).

\pagebreak[4]
\begin{center}
TABLE 3. NGC7099 Stars with $V$, $J$, and $K$ measurements
\end{center}

\begin{center}
\begin{tabular}{rrrr}
\hline\hline
$n^a$ & $K$ & $J-K$ & $V-K$ \\
\hline
  17 &   14.43 &    0.17 &    0.59 \\
  19 &   12.84 &    0.59 &    2.21 \\
  23 &   13.80 &    0.25 &    1.03 \\
  24 &   13.04 &    0.59 &    1.84 \\
  32 &   10.51 &    0.65 &    2.70 \\
  28 &   11.73 &    0.64 &    2.46 \\
  25 &   11.79 &    0.60 &    2.30 \\
  18 &   14.33 &    0.14 &    0.52 \\
   6 &   12.25 &    0.60 &    2.23 \\
   5 &   12.26 &    0.64 &    2.27 \\
   4 &   11.83 &    0.64 &    2.31 \\
  27 &   10.63 &    0.63 &    2.60 \\
  26 &   10.64 &    0.69 &    2.66 \\
  29 &   12.87 &    0.55 &    2.11 \\
  30 &   11.58 &    0.59 &    2.41 \\
  31 &   12.24 &    0.58 &    2.27 \\
  40 &   12.04 &    0.61 &    2.32 \\
  41 &   12.04 &    0.64 &    2.40 \\
  42 &   14.39 &    0.17 &    0.48 \\
  44 &   12.13 &    0.57 &    2.18 \\
  45 &   14.99 &    0.05 &    0.17 \\
  43 &   12.00 &    0.61 &    2.23 \\
  67 &   10.35 &    0.66 &    2.57 \\
  47 &   14.99 &   -0.03 &    0.16 \\
  65 &   13.21 &    0.44 &    1.71 \\
\hline
\end{tabular}
\end{center}

\noindent{$^a$} Identification number from Alcaino \& Liller (1980).

\pagebreak[4]
\begin{center}
TABLE 3. (con't)
\end{center}

\begin{center}
\begin{tabular}{rrrr}
\hline\hline
$n^a$ & $K$ & $J-K$ & $V-K$ \\
\hline
  49 &   13.97 &    0.39 &    1.14 \\
  66 &   12.63 &    0.57 &    2.23 \\
  63 &   12.73 &    0.66 &    2.17 \\
  64 &   10.88 &    0.67 &    2.62 \\
  62 &   10.33 &    0.71 &    2.79 \\
  61 &   12.74 &    0.53 &    2.16 \\
  70 &   12.61 &    0.60 &    2.21 \\
  73 &   11.90 &    0.58 &    2.22 \\
  72 &   12.55 &    0.60 &    2.23 \\
  74 &   12.08 &    0.63 &    2.34 \\
  75 &   11.27 &    0.61 &    2.54 \\
  71 &   12.77 &    0.57 &    2.09 \\
  77 &   10.48 &    0.67 &    2.71 \\
  76 &   14.54 &    0.15 &    0.43 \\
\hline
\end{tabular}
\end{center}

\noindent{$^a$} Identification number from Alcaino \& Liller (1980).

\pagebreak[4]
\begin{center}
TABLE 4. Normal Points
\end{center}

\begin{center}
\begin{tabular}{rcc}
\hline\hline
$K$ & $(J-K)_{5946}$ & $(J-K)_{7099}$ \\
\hline
9.5 & 1.09 & $-$ \\
10.0 & 1.00 & 0.64 \\
10.5 & 1.00 & 0.66 \\
11.0 & 0.94 & 0.63 \\
11.5 & 0.91 & 0.58 \\
12.0 & 0.90 & 0.59 \\
12.5 & 0.87 & 0.57 \\
13.0 & 0.84 & 0.56 \\
13.5 & 0.82 & 0.53 \\
14.0 & 0.79 & 0.51 \\
14.5 & $-$ & 0.50 \\
15.0 & $-$ & 0.48 \\
15.5 & $-$ & 0.45 \\
\hline
\end{tabular}
\end{center}

\pagebreak[4]
\begin{center}
TABLE 5. NGC5946 Color Measurements
\end{center}

\begin{center}
\begin{tabular}{lccc}
\hline\hline
Radius & $(J-K)^a$ & $(J-K)^b$ & Error \\
(arcsec) & & & \\
\hline
0.5 & 0.88 & 0.94 & $\pm 0.04$ \\
1.4 & 0.90 & 0.88 & $\pm 0.05$ \\
2.8 & 0.92 & 0.84 & $\pm 0.05$ \\
5.6 & 0.88 & 0.78 & $\pm 0.06$ \\
11.3 & 0.96 & 0.78 & $\pm 0.08$ \\
22.6 & 0.84 & 0.71 & $\pm 0.30$ \\
\hline
\end{tabular}
\end{center}

\noindent{$^a$} Color from direct aperture measurement.

\noindent{$^b$} Color derived from scheme discussed by Piotto et al. (1988).

\pagebreak[4]
\begin{center}
TABLE 6. NGC7099 Color Measurements
\end{center}

\begin{center}
\begin{tabular}{lccc}
\hline\hline
Radius & $(J-K)^a$ & $(J-K)^b$ & Error \\
(arcsec) & & & \\
\hline
0.4 & 0.39 & 0.34 & $\pm 0.05$ \\
1.1 & 0.46 & 0.36 & $\pm 0.05$ \\
2.2 & 0.45 & 0.40 & $\pm 0.05$ \\
4.4 & 0.53 & 0.38 & $\pm 0.06$ \\
8.9 & 0.49 & 0.41 & $\pm 0.12$ \\
17.8 & 0.55 & 0.43 & $\pm 0.23$ \\
\hline
\end{tabular}
\end{center}

\noindent{$^a$} Color from direct aperture measurement.

\noindent{$^b$} Color derived from scheme discussed by Piotto et al. (1988).

\pagebreak[4]
\begin{center}
TABLE 7. The Brightest Giants in Metal-Poor Clusters
\end{center}

\begin{center}
\begin{tabular}{lrrr}
\hline\hline
Cluster & Star & $K$ & M$_{K}^{Tip}$ \\
\hline
NGC4590 & I-260 & 9.52 & $-5.40$ \\
NGC5024 & IR-1 & 10.61 & $-5.73$ \\
NGC6341 & III-13 & 8.93 & $-5.47$ \\
NGC7078 & I-12 & 9.42 & $-5.56$ \\
\hline
\end{tabular}
\end{center}

\pagebreak[4]
\begin{center}
REFERENCES
\end{center}
\parindent=0.0cm

Alcaino, G. 1978, A\&AS, 33, 185

Alcaino, G., \& Liller, W. 1980, AJ, 85, 1330

Alcaino, G., \& Wamsteker, W. 1982, A\&AS, 50, 141

Alcaino, G., Liller, W., Alvardo, F., \& Wenderoth, E. 1991, AJ, 102, 1371

Armandroff, T. A. 1989, AJ, 97, 375

Armandroff, T. E., \& Zinn, R. 1988, AJ, 96, 92

Bessell, M. S., \& Brett, J. M. 1988, PASP, 100, 1134

Bica, E. L. D., \& Pastoriza, M. G. 1983, Ap. Space Sci., 91, 99

Bolte, M. 1987, ApJ, 319, 760

Bolte, M. 1989, ApJ, 341, 168

Brodie, J. P., \& Hanes, D. A. 1986, ApJ, 300, 258

Brodie, J. P., \& Huchra, D. A. 1990, ApJ, 362, 503

Buonanno, R., Caloi, V., Castellani, V., Corsi, C. E., Ferraro, I., \&
\linebreak[4]\hspace*{1.0cm}Piccolo, F. 1988, A\&AS, 74, 353

Casali, M., \& Hawarden, T. 1992, JCMT-UKIRT Newsletter, 4, 33

Chernoff, D.F., \& Djorgovski, S. 1989, ApJ, 339, 904

Cohen, J. G., \& Sleeper, C. 1995, AJ, 109, 242

Cohen, J. G., Frogel, J. A., \& Persson, S. E. 1978, ApJ, 222, 165

Cordoni, J.-P., \& Auriere, M. 1984, A\&AS, 58, 559

Davidge, T. J. 1992, AJ, 103, 1259

Davidge, T. J., \& Simons, D. A. 1994, ApJ, 423, 640

Depoy, D. L., Atwood, B., Byard, P., Frogel, J., \& O'Brien, T. 1993, SPIE,
\linebreak[4]\hspace*{1.0cm}1946, 667

Dickens, R. F. 1972, MNRAS, 157, 299

Djorgovski, S., \& Piotto, G. 1992, in Structure and Dynamics of Globular
\linebreak[4]\hspace*{1.0cm}Clusters, ASP Conference Series \# 50, ed. S. G.
Djorgovski \linebreak[4]\hspace*{1.0cm}\& G. Meylan, pp. 203

Djorgovski, S., Piotto, G., Phinney, E. S., \& Chernoff, D. F. 1991, ApJ,
\linebreak[4]\hspace*{1.0cm}372, L41

Elias, J. H., Frogel, J. A., Matthews, K., \& Neugebauer, G. 1982, AJ, 87,
\linebreak[4]\hspace*{1.0cm}1029

Frogel, J. A. 1983, ApJ, 272, 167

Frogel, J. A., Cohen, J. G., \& Persson, S. E. 1983, ApJ, 275, 773

Frogel, J. A., Persson, S. E., \& Cohen, J. G. 1981, ApJ, 246, 842

Frogel, J. A., Persson, S. E., \& Cohen, J. G. 1983, ApJS, 53, 713

Geisler, D., Minniti, D., \& Claria, J. J. 1992, AJ, 104, 627

Hesser, J. E., \& Shawl, S. J. 1986, PASP, 97, 465

Hodapp, K.-W., Rayner, J., \& Irwin, E. 1992, PASP, 104, 441

Liller, M. H. 1983, AJ, 88, 404

Minniti, D., Geisler, D., Peterson, R. C., \& Claria, J. J. 1993, ApJ, 413,
\linebreak[4]\hspace*{1.0cm}548

Pastoriza, M. G., Bica, E. L. D., Copetti, M. V. F., \& Dottori, H. A. 1986,
\linebreak[4]\hspace*{1.0cm}Astr. Space Sci., 119, 279

Piotto, G., King, I. R., \& Djorgovski, S. 1988, AJ, 96, 1918

Piotto, G., Capacioli, M., Ortolani, S., Rosino, L., Alcaino, G., Liller,
\linebreak[4]\hspace*{1.0cm}W. 1987, AJ, 94, 360

Piotto, G., King, I. R., Capaccioli, M., Ortolani, S., \& Djorgovski, S.
\linebreak[4]\hspace*{1.0cm}1990, ApJ, 350, 662

Racine, R., \& Harris, W. E. 1989, AJ, 98, 1609

Richer, H. B., Fahlman, G. G., \& VandenBerg, D. A. 1988, ApJ, 329, 187

Rieke, G. H., \& Lebofsky, M. J. 1985, ApJ, 288, 618

Shapley, H. 1918, ApJ, 48, 154

Stetson, P. B. 1987, PASP, 99, 191

Stetson, P. B., \& Harris, W. E. 1988, AJ, 96, 909

Trager, S. C., King, I. R., \& Djorgovski, S. 1995, AJ, 109, 218

Wainscoat, R. J., \& Cowie, L. L. 1992, AJ, 103, 332

Yanny, B., Guhathakurta, P., Schneider, D. P., \& Bahcall, J. N. 1994, ApJ,
\linebreak[4]\hspace*{1.0cm}435, L59

Zinn, R., \& West, M. J. 1984, ApJS, 55, 45

\pagebreak[4]
\begin{center}
FIGURE CAPTIONS
\end{center}

FIG. 1. Final $K$ image of NGC5946 Field 1. The dimensions of this image
are roughly $90 \times 93$ arcsec.

\vspace{0.5cm}
FIG. 2. Final $K$ image of NGC5946 Field 2. The dimensions of this image
are roughly $83 \times 81$ arcsec.

\vspace{0.5cm}
FIG. 3. Final $K'$ image of NGC7099 field. The image dimensions are
roughly $75 \times 94$ arcsec.

\vspace{0.5cm}
FIG. 4. $(K, J-K)$ CMD's of, from left to right, NGC5946 Field 1, NGC5946
Field 2, and NGC7099. Only stars with errors from ALLSTAR $\leq 0.07$ mag have
been included in this figure.

\vspace{0.5cm}
FIG. 5. $(K, V-K)$ CMD's of NGC5946 Field 1 (left hand side) and NGC7099
(right hand side). The $V$ measurements are those given by Alcaino
et al. (1991 $-$ NGC5946) and Alcaino \& Liller (1980 $-$ NGC7099).

\vspace{0.5cm}
FIG. 6. The $J-H, H-K$ two-color diagram for stars in NGC5946 Field 1 with
$K \leq 13$. Also shown are bright stars in M13 (open squares) and
M92 (filled squares) from Cohen et al. (1978), shifted to match the reddening
of NGC5946. The solid line is the locus of solar neighborhood giants from
Bessell \& Brett (1988), also shifted to match the reddening of NGC5946.

\vspace{0.5cm}
FIG. 7. The $J-K, V-K$ two-color diagram for stars in NGC5946 (top panel)
and NGC7099 (bottom panel). Also shown are bright giants in M13 (open squares)
and M92 (filled squares) from Cohen et al. (1978), as well as the locus of
solar neighborhood giants from Bessell \& Brett (1988 $-$ solid line). The M13,
M92, and solar neightborhood data have been shifted to match the reddenings of
NGC5946 and NGC7099.

\vspace{0.5cm}
FIG. 8. The $J-K$ color profiles for NGC5946 (open squares) and NGC7099
(filled squares). Colors resulting from direct aperture measurements
are shown in the top panel, while those made using the segmental median
technique of Piotto et al. (1988 $-$ see text) are shown in the lower panel.

\vspace{0.5cm}
FIG. 9. NGC5946 Field 1 normal points (crosses) compared with [Fe/H] $= -2$
(left hand panel) and [Fe/H] $= -1.3$ cluster sequences (right hand panel).

\vspace{0.5cm}
FIG. 10. NGC7099 normal points (crosses) compared with [Fe/H] $= -2$
(left hand panel) and [Fe/H] $= -1.3$ cluster sequences (right hand panel).

\vspace{0.5cm}
FIG. 11. CMD's for stars in NGC5946 and NGC7099 falling within various
distances from the cluster centers. The CMD's for NGC5946 sample the
intervals $0 - 9$ (upper left hand corner) and $9 - 16.5$ (lower left
hand corner) arcsec. The CMD's for NGC7099 sample the regions $0 - 10$
(upper right hand corner) and $10 - 19$ (lower right hand corner) arcsec.
\end{document}